\documentclass[journal=apchd5,manuscript=article]{achemso}

\usepackage[version=3]{mhchem} % Formula subscripts using \ce{}
\usepackage[T1]{fontenc}
\usepackage{color}
\author{Adriano Cacciola}
\author{Claudia Triolo}
\author{Omar Di Stefano}
\affiliation[Universit\`{a} di Messina]
{Dipartimento di Fisica e di Scienze della Terra, Universit\`{a} di Messina, Viale F. Stagno d'Alcontres 31, I-98166 Messina, Italy}
\author{Armando Genco}
\affiliation[Istituto Italiano di Tecnologia]
{CBN, Istituto Italiano di Tecnologia, Via Barsanti 1, 73010 Lecce}
\alsoaffiliation[Universit\`{a} del Salento]
{Dipartimento di Matematica e Fisica ``Ennio De Giorgi", Universit\`{a} del Salento, Via per Arnesano, 73100 Lecce, Italy}
\author{Marco Mazzeo}
\affiliation[Universit\`{a} del Salento]
{Dipartimento di Matematica e Fisica ``Ennio De Giorgi", Universit\`{a} del Salento, Via per Arnesano, 73100 Lecce, Italy}
\alsoaffiliation[CNR NANOTEC]
{CNR NANOTEC - Istituto di Nanotecnologia, Polo di Nanotecnologia c/o Campus Ecotekne, via Monteroni - 73100 Lecce, Italy}
\author{Rosalba Saija}
\author{Salvatore Patan\`{e}}
\author{Salvatore Savasta}
\affiliation[Universit\`{a} di Messina]
{Dipartimento di Fisica e di Scienze della Terra, Universit\`{a} di Messina, Viale F. Stagno d'Alcontres 31, I-98166 Messina, Italy}
\email{ssavasta@unime.it}
\title[An \textsf{achemso} demo]
  {Subdiffraction Light Concentration\\ by J-Aggregate Nanostructures}

\begin{document}

\begin{tocentry}

%\begin{figure}
\includegraphics[height=35 mm]{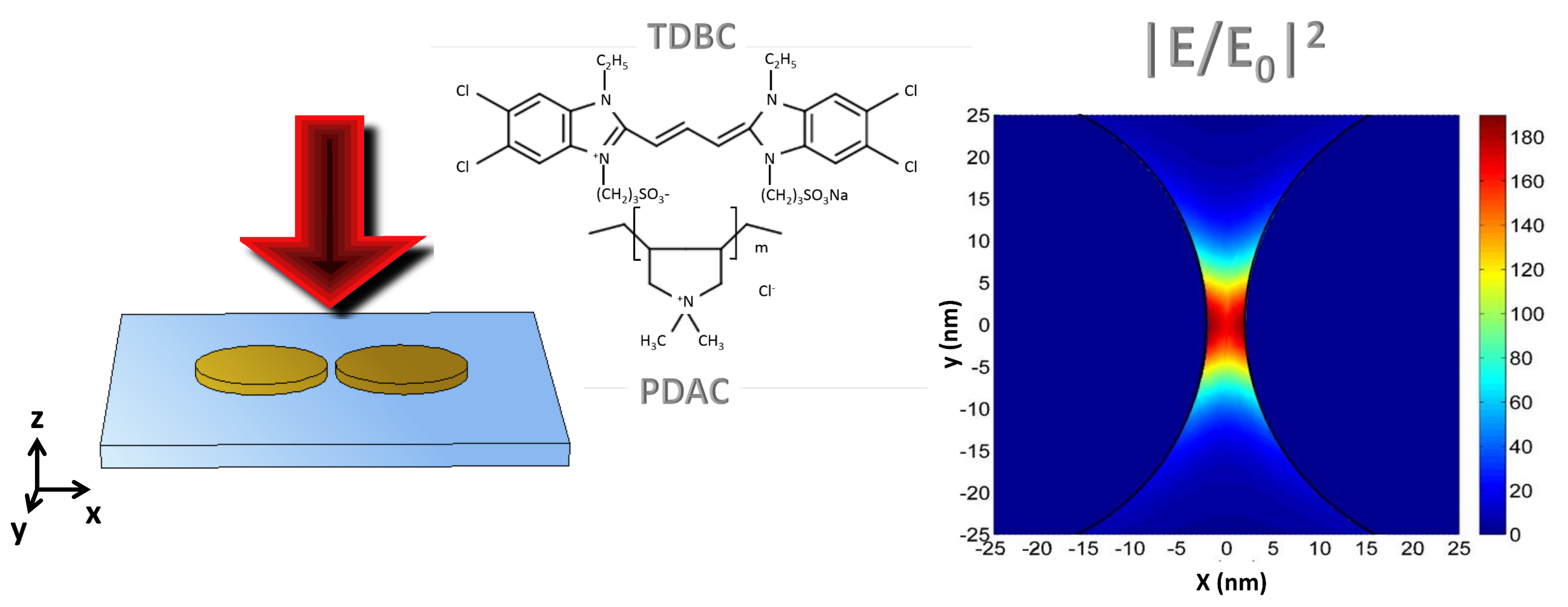}    
%\end{figure}  
%
%Some journals require a graphical entry for the Table of Contents.
%This should be laid out ``print ready'' so that the sizing of the
%Inside the \texttt{tocentry} environment, the font used is Helvetica
%8\,pt, as required by \emph{Journal of the American Chemical
%The surrounding frame is 9\,cm by 3.5\,cm, which is the maximum
%permitted for  \emph{Journal of the American Chemical Society}
%graphical table of content entries. The box will not resize if the
%content is too big: instead it will overflow the edge of the box.
%
%This box and the associated title will always be printed on a
%separate page at the end of the document.

\end{tocentry}

\begin{abstract}
We show, by accurate scattering calculations, that nanostructures obtained from thin films of J-aggregate dyes, despite their insulating behavior, are able to concentrate the electromagnetic field at optical frequencies like metallic nanoparticles. These results promise to widely enlarge the range of plasmonic materials, thus opening new perspectives in nanophotonics. Specifically we investigate ultrathin nanodisks and nanodisk dimers which can be obtained by standard nanolitography and nanopatterning techniques. These molecular aggregates display highly attractive nonlinear optical properties which can be exploited for the realization of ultracompact devices for switching light by light on the nanoscale without the need of additional nonlinear materials. 
\end{abstract}
%\maketitle 



%-------------------------------------------------
%\cite{Maier2001, Gramotnev2010, Schuller2010,Anker2008,Atwater2010}
%======================================
%=======================================
%
%The exceptional ability of metallic nanostructures to concentrate light into volumes far below the diffraction limit has pushed their use in a vast range of nanophotonics technologies and research fields.
%\cite{Maier2001,Schuller2010,Anker2008,Atwater2010}.
%\cite{Wuerthner2011} 

When light interacts with metal nanoparticles and  nanostructures, it can excite collective oscillations known as localized surface plasmons (LSPs) which provide the opportunity to confine light to very small dimensions below the diffraction limit \cite{Maier2001,Maier2007,Gramotnev2010,Schuller2010}.
This high confinement can lead to a striking near-field enhancement which  can significantly enhance weak nonlinear
processes \cite{Kauranen2012} and enables a great variety of applications such as optical sensing \cite{Anker2008,Stewart2008}, higher efficiency solar cells \cite{Atwater2010}, nanophotonics\cite{Maier2001, Willets2007,Kauranen2012} including ultracompact lasers and amplifiers \cite{Noginov2009} and antennas transmitting and receiving light signals at the nanoscale \cite{Schnell2009,Schuller2010,Aouani2011}. The small mode volume of LSP resonances also increases the photonic local density of states (LDOS) close to a plasmonic nanoparticle, enabling the modification of the optical properties (decay rate and quantum efficiency) of emitters placed in its close proximity (see {\em e.g.} Ref.\ \cite{Russell2012} and Supporting Information Figure S1). The interaction of quantum emitters, as
quantum dots or dye molecules, with individual metallic nanostructures carries significant potential for the quantum control of light at the nanoscale   \cite{Maier2001,Ozbay2006,Chang2007, Ridolfo2010, Savasta2010,Zengin2013, Tame2013,Cacciola2014,Antosiewicz2014,Zengin2015}.
As first highlighted by Takahara {\em et al.}\cite{Takahara1997}, only materials with a negative real part of the dielectric function and moderate losses, are able to excite localized surface plasmons and hence to confine light to very small dimensions below the diffraction limit.
These collective and confined excitations are efficiently supported, despite dissipative losses, by noble metals where the
effective response of the electrons can be described by a Drude-Lorentz dielectric function whose real part is negative for frequencies below the plasma frequency \cite{Maier2007}. Also superconductors or graphene have been proposed as a platform for surface plasmon polaritons \cite{Tassin2012,Woessner2014}.

Collective oscillations of free electrons are not the only way a
negative permittivity may arise. It may also occurs in the
high energy tail of a strong absorption resonance. 
For example, it has been shown that lattice vibrations in polar dielectric materials can also give origin to negative dielectric permettivity in the far or mid-infrared spectral range which can support phonon-polaritons confined to the surface \cite{Shchegrov2000,Ocelic2004}.
It has also been shown that in the near-infrared and optical spectral ranges, sharp excitonic resonances in organic molecules can give origin to spectral regions where the dielectric permettivity is negative \cite{Philpott1979,Gu2013a}.
The response of Lorentzian/excitonic nanoshells with optically inactive core and embedding medium
have been also studied in the quasistatic limit and compared  with the properties of the Drude/plasmonics nanoshells \cite{Gulen2013}.
Recently, it has been theoretically shown that a nanosphere of TDBC-doped PVA displays enhanced
optical fields and subwavelength field confinement analogously to plasmonic nanoparticles \cite{Gentile2014}. 
At present, however, spherical nanoparticles of J-aggregates have not been realized and the synthesis of high quality nanospheres does not seem readily achievable.

Here we propose more realistic J-aggregates nanostructures suitable for obtaining on chip strong subwavelength light confinement.
%We explore the potential of supremolecular organic nanostructures to concentrate light at dimensions below the diffraction limit.
By accurate scattering calculations, we show that thin films of J-aggregate dyes \cite{Bradley2005} can be exploited for the realization of plasmon-like nanostructures working at optical frequencies. Such nanostructures can be realized after
the film growth by standard nanolitography and nanopatterning techniques. The proposed
structures consist of ultrathin nanodisks on a silica substrate. Moreover, we show that these molecular aggregates can also be exploited for the realization of more complex plasmon-like nanostructures to achive very high field concentration. Specifically we find that nanodisk dimers with radius $a = 30$ nm are able to generate near-field enhancements $|E/E_0|^2$ beyond 350.
The here proposed J-aggregate nanostructures are promising towards the emerging of a low-cost plasmonic technology for the development of ultracompact photonic and optoelectronic devices.
Moreover, J-aggregates display significant optical nonlinearities \cite{Minoshima1994,Vasa2013} which can induce switchable localized surface resonances (LSRs) required for the realization of plasmonic devices \cite{Kauranen2012} without the need to exploit the coupling with additional nonlinear optical materials \cite{Fofang2011,Vasa2013}. This target has not been achieved by  metallic nanoparticles owing to their intrinsically weak optical nonlinearities which can be overcome only by coupling LSPs to nonlinear optical resonances \cite{Fofang2011,Vasa2013}.

 J-aggregates self-assembled nanostructures have been realized in various forms including quasi-one dimensional chains, single monolayers, tubes, etc. which could be employed as molecular plasmonic materials \cite{Saikin2013}. One limitation of these surface resonances is that they can manifest in a limited spectral region in the high energy side of the excitonic resonance. However,  because of the huge variety of organic dyes that can aggregate, it could be possible  to create J-aggregates LSRs within an arbitrary part of visible and near IR spectrum \cite{Walker2011} and there is room for chemically engineering and optimizing their plasmonic behavior in order to minimize losses. Moreover J-aggregates emit light, generally with good quantum efficiencies, thus their LSRs could be exploited for the realization of efficient light-emitting nanoantennas. 
%=====================================================
\begin{figure}
  \includegraphics[height= 60 mm]{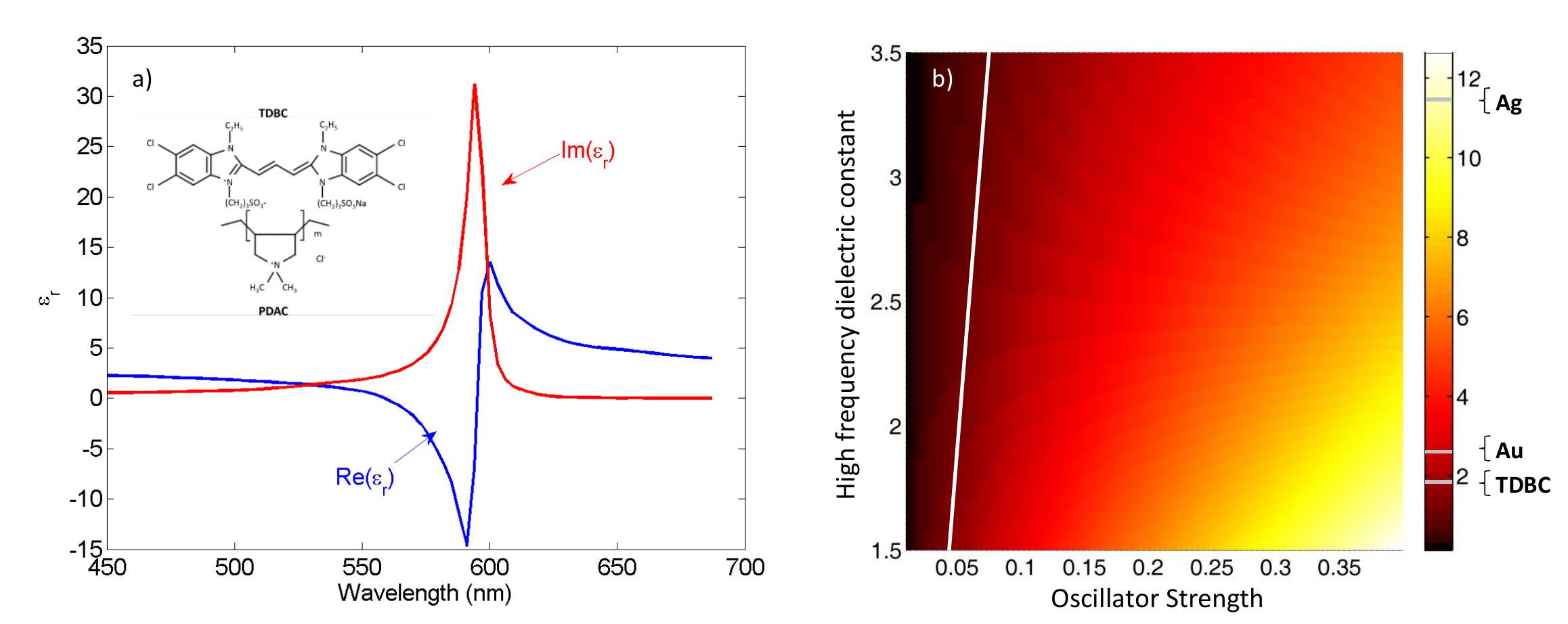}
  \caption{Plasmon-like properties in the blue-side spectral region of strong excitonic resonances. (a) Real and imaginary part of the dielectric function directly derived by the complex refractive index ($\tilde n = n + i \kappa$) obtained via a Kramers-Kronig regression of reflectance data \cite{Bradley2005} of layer by layer assembled TDBC/PDAC thin films. The real part of the dielectric function is negative in the blue-side region, a key feature for confining light below the diffraction limit.
	The inset shows the molecular structure of the polycation PDAC and of the anyon TDBC, a J-aggregate forming cyanine dye. (b)
Colormap of the peak extinction efficiency $Q_{\rm ext} = \sigma_{\rm ext}/(\pi a^2)$ calculated in the quasistatic approximation for a 40 nm sphere in vacuum by using the Lorentz dielectric function as a function of the normalized oscillator strength $f$ and the high-frequency dielectric constant $\varepsilon_{\infty}$ with fixed $Q_0 = 80$.
The white line in the contourplot separates the region where $\varepsilon_{\rm min} < -2$ (on the right side) and plasmon-like effects starts to be expected, from the left region  where  $\varepsilon_{\rm min} > -2$. This model map points out the plasmonic potentiality of excitonic resonances with high oscillator strengths. The corresponding values of $Q_{\rm ext}$ calculated for Ag, Au, and TDBC/PDAC nanoparticles are indicated in the color-bar for reference.}
	%\label{lev}
\end{figure}
%=====================================================
%These systems have attracted a lot of attention, in particular in view of their optical properties\cite{Wuerthner2011}.

These J-aggregates possess remarkable morphological and optical properties that make them suited for use in photonic and optoelectronic applications \cite{Lidzey1998, Tischler2005, Bellessa2004,Schlather2013}.
The broad absorption line corresponding to the single molecule excitation is shifted to the red side of the spectrum and transforms into a collective narrow-linewidth optical transition possessing a giant oscillator strength derived from the coherent coupling of the aggregated molecules \cite{Wuerthner2011}.
%Aggregates composed of TDBC fluorescent molecules have been produced in solution and on surfaces.
Layer by layer growth of cyanine-dyes (TDBC) aggregates to form an ordered thin film has been demonstrated with a peak absorption constant of $\alpha \simeq 1 \times 10^6$ cm$^{-1}$, one of the highest reported for a thin film\cite{Bradley2005}. 
The Frenkel excitons of these high absorptive materials have been exploited in demonstrations of strong coupling between excitons and microcavity photons \cite{Lidzey1998},  room-temperature operation of exciton-optoelectronic devices \cite{Tischler2005}, strong coupling between excitons and surface plasmon-polaritons \cite{Bellessa2004} or LSP resonances \cite{Schlather2013}.

The dielectric function at frequencies around sharp excitonic resonances is usually quite well reproduced by the Lorentz model: $\varepsilon(\omega) = \varepsilon_\infty +  \frac{f \omega_0^2}{\omega_0^2-\omega^2 -i \gamma_0 \omega}$, where $\omega_0$ is the resonance frequency, $\gamma_0$ the width of the resonance, $f$ the dimensionless oscillator strength and $\varepsilon_\infty$ the high-frequency background constant. In contrast to the Drude dielectric function, whose real part is negative for frequencies below the plasma frequency, the real part of the Lorentz dielectric function is generally positive for almost all excitonic absorption lines.  It takes its minimum $\varepsilon'_{\rm min} \simeq \varepsilon_\infty - f \omega_0 /(2 \gamma_0)$ at  $\omega_{\rm min} \simeq \omega_0 + \gamma_0/2$.
However there is no specific constraint preventing negative values.
For narrow absorption lines and high oscillator strengths,  the real part of the Lorentz dielectric function can acquire negative values on the high energy side of the excitonic resonance if $f Q_0 /(2 \varepsilon_\infty) > 1$, where $Q_0 = \omega_0/\gamma_0$ is the quality factor of the excitonic line.
%However at $\omega_{\rm min}$ losses can be very high, especially for high oscillator strengths, since the imaginary part of the dielectric function %$\varepsilon''(\omega_{\rm min})$ is half of the peak value at $\omega = \omega_0$. 
The giant oscillator strength and the strong spectral narrowing of TDBC aggregates, both induced by the coherent intermolecular coupling, result in ordered thin films with a negative dielectric function in the high energy tail of the excitonic resonance with an astonishingly low minimum value $\varepsilon_{\rm min} \simeq -14$. Figure 1a displays the dielectric function directly derived by the the real and imaginary components of the refractive index ($\tilde n = n + i \kappa$) obtained via a Kramers-Kronig regression of neat films reflectance data \cite{Bradley2005}. The obtained optical constants accurately reproduce both the reflectance and transmittance data for a $5$ nm thick thin film and the reflectance of more complex planar optical structures containing a thin TDBC/PDAC film \cite{Tischler2006}.

 The obtained TDBC/PDAC dielectric function shown in Fig.\ 1a qualitatively resembles a Lorentz dielectric function.
However its imaginary part displays a tail in the blue part of the spectrum due to the presence of the vibrational degrees of freedom and the Kramers-Kronig related real part is also asymmetric.
%differs from the ideal Lorentz model, it can be approximately described by it. A fit gives $f \simeq 0.4$, $\varepsilon_\infty = 2.31$, $\gamma_0 \simeq 28$ meV, and %$Q_0 \simeq 72$. 
%Figure 1a also shows that at $\omega_{\rm min}$, the imaginary part of the dielectric function is very high causing strong losses. However, for increasing frequencies, the decay towards zero of the imaginary part $\varepsilon''(\omega)$ is faster than the decay  towards $\varepsilon_\infty$ of $\varepsilon'(\omega)$, so that there is an interesting spectral region where losses are moderate and $\varepsilon'(\omega)$ is still negative.

\section{Results}
%LSP resonances depend on the particle size, shape and refractive index of their surroundings \cite{Willets2007}. Moreover in the presence of more %than one particle, their spectrum as well as the resulting field enhancement strongly depends on the distance between them \cite{Halas2011}.
In order to explore the potentialities of J-aggregates as plasmonic materials, we start with the most convenient geometry for an analytical treatment: a spherical nanoparticle of radius $a$ much smaller than the effective wavelength $\lambda/\sqrt{\varepsilon_{\rm d}}$, where $\varepsilon_{\rm d}$ is the dielectric constant (assumed to be real) of the surrounding medium. Scattering calculations on more feasible J-aggregates nanostructures  will be presented later.
The signature of LSP effects can be easily recognized by the polarizability $\alpha_{\rm p} = P/(\varepsilon_0 \varepsilon_{\rm d}E_0)$, defined as the ratio between the dipole moment $P$ induced in the nanoparticle by an incoming wave and the amplitude of the incident displacement field $E_0$. In the quasi-static approximation, for a spherical nanoparticle,
$\alpha_{\rm p} = 4 \pi a^3 (\varepsilon (\omega) - \varepsilon_{\rm d})/(\varepsilon (\omega) + 2 \varepsilon_{\rm d})$. The nanoparticle polarizability displays a dipole resonance when $\varepsilon' (\omega)= -2 \varepsilon_{\rm d}$ known as Fr\"{o}lich condition. The optical theorem provides a direct link between the polarizability and the extinction cross section $\sigma_{\rm ext} = \sqrt{\varepsilon_{\rm d}}\,  k_0\,  {\rm Im}\{ \alpha_{\rm p} \}$, where $k_0 = 2\pi/\lambda$.
This resonance also produces a resonant enhancement of the dipolar field and of the optical LDOS around the nanoparticle. 
Many of the attractive plasmonic applications of metal nanoparticles and nanostructures rely on this field-enhancement at the plasmon resonance \cite{Maier2001, Gramotnev2010, Schuller2010,Ozbay2006,Willets2007,Anker2008,Atwater2010}.
For a TDBC nanoparticle in vacuum, the dipole LSR occurs at  $\lambda \simeq 570$ nm where $\varepsilon = -2 + i 3.5$. 
In Fig.\ 1b a model-map points out the plasmonic potentiality of strong excitonic resonances. The figure displays the peak extinction efficiency $Q_{\rm ext} = \sigma_{\rm ext}/(\pi a^2)$ for a 40 nm sphere calculated within the quasistatic approximation by using the Lorentz dielectric function versus the normalized oscillator strength $f$ and the high-frequency dielectric constant $\varepsilon_{\infty}$ with fixed $Q_0 = 80$. 
%=====================================================
\begin{figure}
 \includegraphics[height= 120 mm]{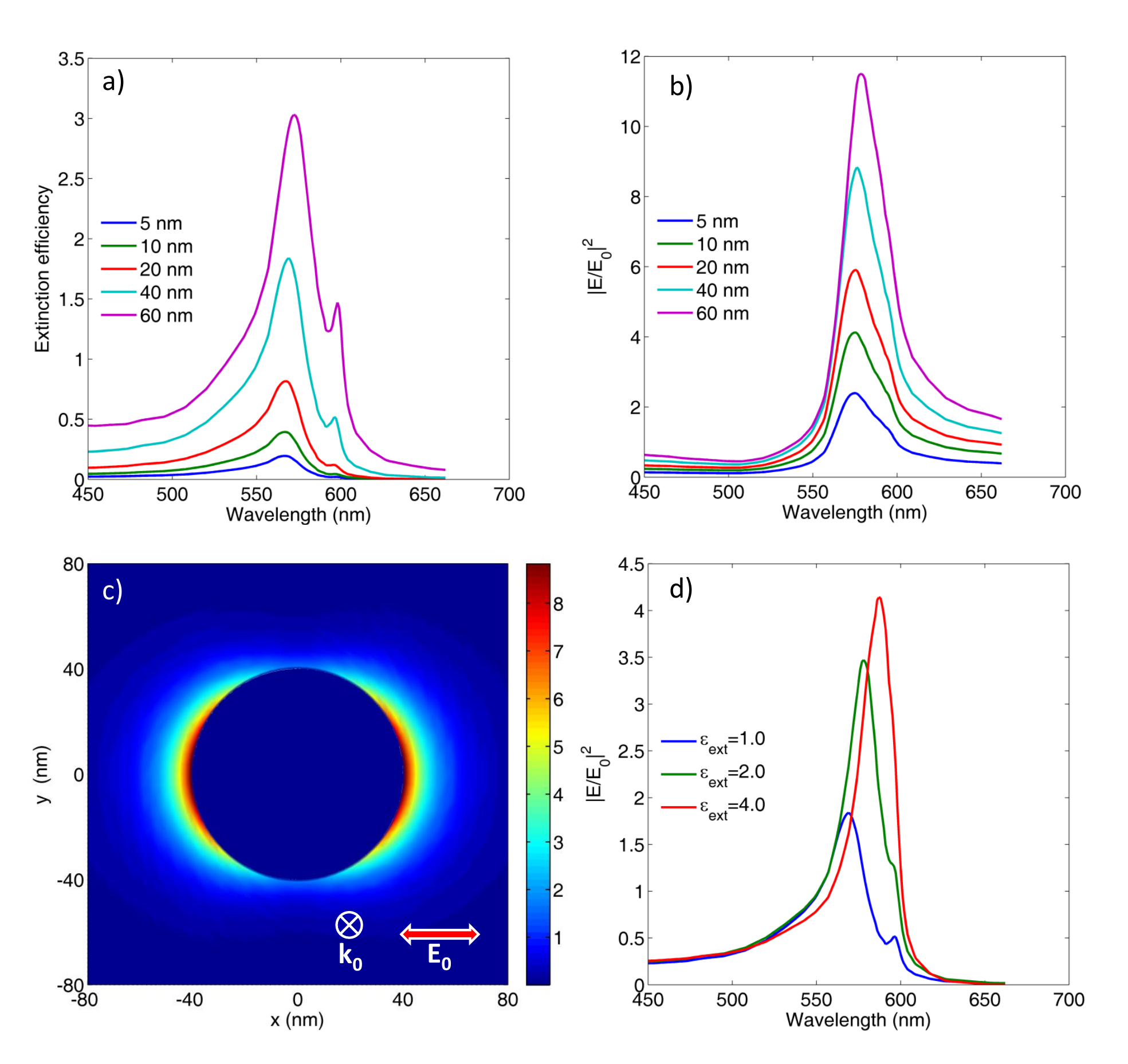}
  \caption{Scattering calculations on TDBC/PDAC nanospheres. (a) Extinction efficiency $Q_{\rm ext}$ for TDBC spherical nanoparticles in vacuum of different radii $a$, calculated beyond the quasistatic approximation, by employing the Mie theory implemented within the T-matrix formalism. (b) Near-field enhancement spectra $|E/E_0|^2$ for the TDBC spherical nanoparticles. (c) Near-field enhancement distribution on the plane containing the polarization axis and normal to the propagation direction for a  TDBC nanoparticle with a radius $a = 40$ nm illuminated by a plane-wave  at $\lambda = 578$ nm.
(d) Near-field enhancement spectra as in (b) but with different dielectric constants of the external surrounding medium.}\label{2}
\end{figure}
%=====================================================

Figure 2a displays the extinction efficiency $Q_{\rm ext}$ for TDBC spherical nanoparticles of different radii $a$, calculated beyond the quasistatic approximation, by employing the Mie theory implemented within the T-matrix formalism\cite{Borghese2007,Cacciola2014} (see also the Method Sect.).
It is remarkable that a $Q_{\rm ext}$ slightly larger than 3 is observed for a nanoparticle with a radius of 60 nm. This means that the interception area that
the spherical particle presents to the incident radiation is about three times larger than its actual size. Such extinction efficiency is lower than that for a gold nanoparticle with the same radius (see also the colorbar in Fig. 1b), but it is significantly larger than the efficiency  of dielectric particles with positive dielectric functions. It is worth noticing that the extinction spectra displays peaks at the wavelength corresponding to the LSR which is significantly blue-shifted with respect to the excitonic absorption maximum (compare Fig.\ 1 and Fig.\ 2a). Figure 2a also displays a significantly smaller peak corresponding to the excitonic resonance.    
Figure 2b shows the calculated near-field enhancement spectra $|E/E_0|^2$ for the TDBC nanoparticles. The field has been calculated at a distance $d =1$ nm from the surface of the nanoparticle at an angle between the incident electric field and the position vector (with the origin at the nanoparticle center) $\theta =0$. A significant plasmon-like near-field enhancement (beyond one order of magnitude for the $a = 60$ nm nanoparticle)  can be observed. Such enhancement is about one half of the enhancement induced by a gold nanoparticle of the same size. The wavelength where the maximum enhancement occurs is red-shifted with respect to the corresponding  extinction-spectrum peak, which is a typical plasmonic feature \cite{Zuloaga2011}.
Figure 2c shows the  near-field enhancement distribution on the plane containing the polarization axis and normal to the propagation direction for a  TDBC nanoparticle with a radius $a = 40$ nm. Calculations have been performed considering an incident plane-wave  at $\lambda = 578$ nm. The influence of the dielectric environment on the LSR resonance is shown in Fig.\ 2d. A shift of the resonance due to the modification of the Fr\"{o}lich condition can be observed.
By confining light using LSPs, it is possible to significantly alter the  photonic LDOS, thus enabling a remarkable enhancement of the spontaneous emission rate of emitters placed in the near-field of a metallic nanoparticle. For an emitter oriented towards a $40$ nm TDBC nanoparticle center, located at a distance $d= 5$ nm with a transition wavelength  $\lambda_0 = 578$ nm, we find a spontaneous emission enhancement $\Gamma/\Gamma_0 \simeq 10$ (Supporting Information Figure SI1). 
%=====================================================
\begin{figure}
  \includegraphics[height= 120 mm]{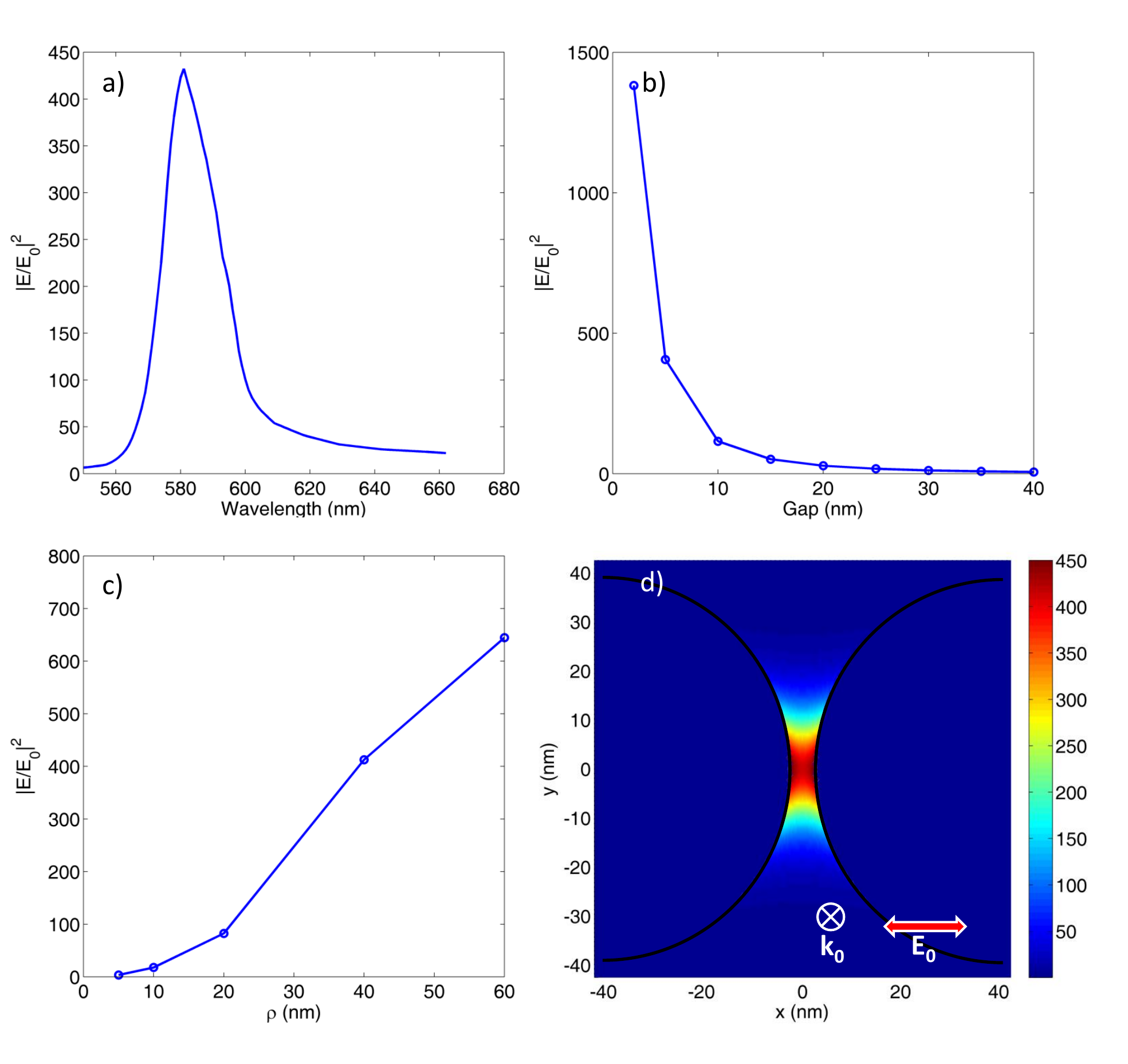}
  \caption{Scattering calculations for a dimer of two nanospheres in vacuum illuminated by longitudinally polarized plane waves. (a) Near-field enhancement spectrum calculated at the center of a 5 nm gap between two nanospheres (radius $a = 40$ nm). (b) Peak field enhancement at the center of the gap as a function of the gap ($a = 40$ nm). (c) Peak field-enhancement at the center of a 5 nm gap as a function of the nanoparticles radius. (d) Near-field enhancement distribution around the nanostructure on the plane z = h/2 for h=5 nm calculated for an incident plane wave at $\lambda$= 581 nm.
	}\label{3}
\end{figure}
%=====================================================

When two or more plasmon-resonant nanoparticles are closely spaced, LSPs of individual particles interact, giving rise to a strong optical field enhancement in the gap regions between the particles \cite{Nordlander2004,Talley2005, Halas2011}. 
As a further signature of the plasmonic behaviour of TDBC nanoparticles, we show that such a strong enhancement can be observed with J-aggregate TDBC dimers. The near-field enhancement spectrum $|E/E_0|^2$ calculated at the center of the gap between two nanospheres with $a = 40$ nm  is reported in Fig.\ 3a. All the presented calculations for dimers have been obtained considering a longitudinally polarized incident field (along the dimer axis).
The peak field enhancement at the center of the gap between two TDBC spherical nanoparticles ($a = 40$ nm) as a function of the gap is reported in Fig.\ 3b. Figure 3c shows the peak field-enhancement at the center of a 5 nm gap as a function of the nanoparticles radius. The  near-field enhancement distribution on the plane containing the dimer axis and orthogonal to the propagation direction  is displayed in Fig.\ 3d.
At present, spherical nanoparticles of J-aggregates have not been realized and we are not aware of any present growing technique. Moreover, the dielectric function here considered has been derived by means of optical measurements on an ordered TDBC thin film grown layer by layer on a substrate. 

We thus study more feasible J-aggregates nanostructures, showing a plasmon-like behavior analogous to the above described nano-spheres.
Such nanostructures can be realized after the film growth by standard nanolitography and nanopatterning techniques \cite{Menard2007}.
The proposed structures consist of ultrathin nanodisks and nanodisk dimers grown on a silica substrate.
Scattering calculations have been performed by the finite element method (see Methods). 
%=====================================================
\begin{figure}
  \includegraphics[height= 80 mm]{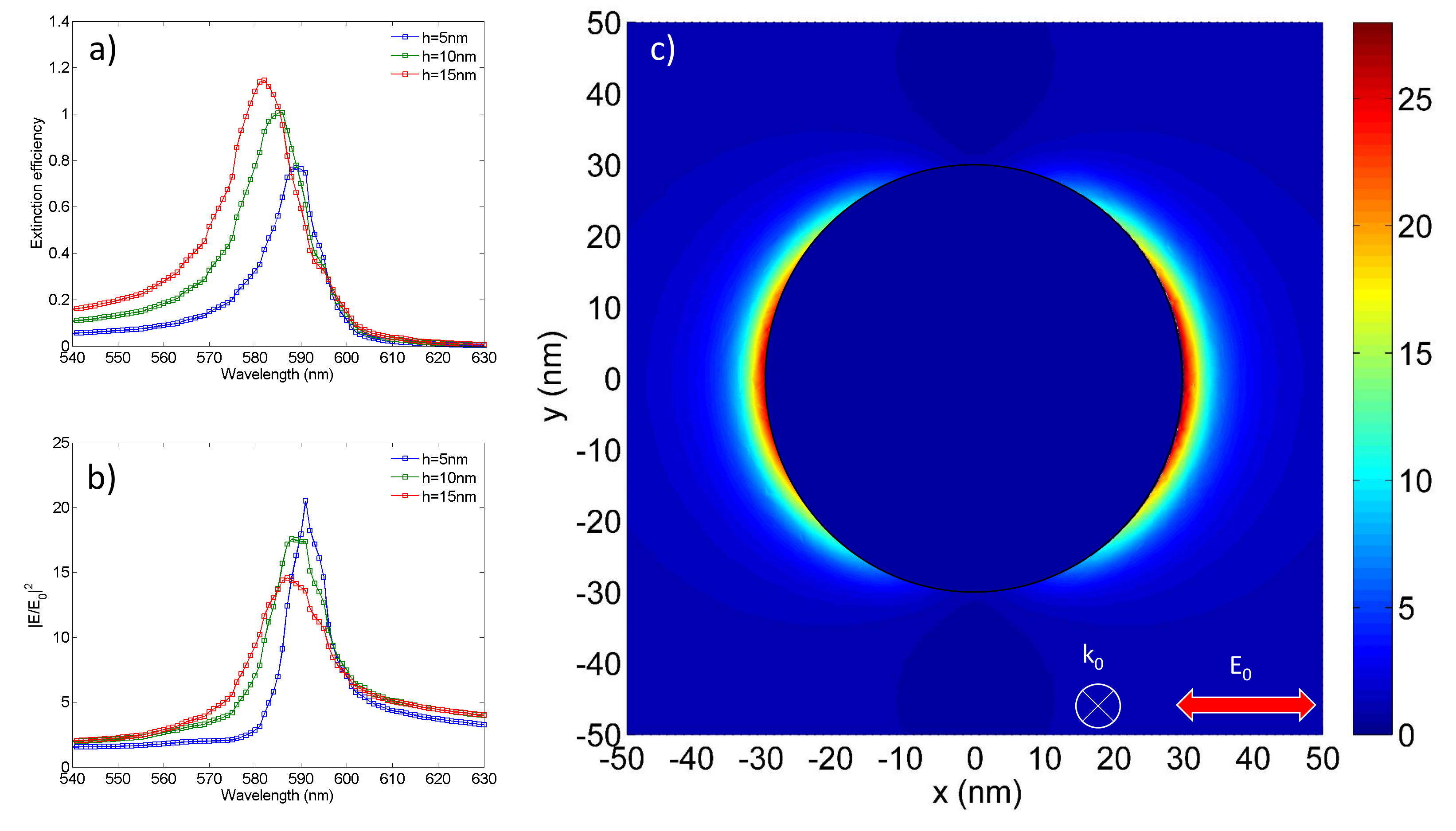}
  \caption{Scattering calculations on single nanodisks. (a) Extinction efficiency $Q_{\rm ext}$ spectra for three single nanodisks  of radius $a = 30$ nm and heights $h= 5$, 10, 15 nm, calculated under normal illumination. (b) Near-field enhancement calculated at $z = h/2$ and at a distance $d = 1$nm from the disk lateral surface along the polarization direction ($\theta =0$). (c) Near-field enhancement distribution on the plane $z = h/2$ and normal to the propagation direction for the TDBC nanodisk  with $h = 5$ nm, under plane wave illumination at $\lambda = 591$ nm. 
	}\label{4}
\end{figure}
%=====================================================
Figure 4a displays the extinction efficiency $Q_{\rm ext}$ spectra for three single nanodisks  of radius $a = 30$ nm and heights $h= 5$, 10, 15 nm, calculated under normal illumination. Increasing the thickness $h$, a blue-shift of the LSR is observed. 
This blue-shift at increasing ratios $h/a$ is a typical plasmonic feature reported for metal nanodisks \cite{Zoric2011}.
The small shoulder on the red side of the peak is a signature of the exciton absorption peak at about $\lambda = 595$ nm corresponding to the wavelength where the imaginary part of the TDBC/PDAC dielectric function is maximum (see Fig.\ 1a). 
The near-field enhancement $|E/E_0|^2$ calculated at an angle between the incident electric field and the position vector (with the origin at the nanodisks center) $\theta =0$, at $z = h/2$, and at a distance $d = 1$nm from the disk lateral surface, is shown in Fig.\ 4b. 
Figure 4c displays the  near-field enhancement distribution on the plane containing the polarization axis and normal to the propagation direction  for the TDBC nanodisk  with $h = 5$ nm. Calculations have been performed considering an incident plane-wave  at the near-field enhancement peak in Fig.\ 4b. A strong field enhancement beyond 25 is observed. These results clearly show that realistic J-aggregate nanostructures are able to support LSP-like resonances. For comparison, scattering calculations for gold nanodisks have been included in the Supporting Information (Fig.\ SI2).
%=====================================================
\begin{figure}
  \includegraphics[height= 120 mm]{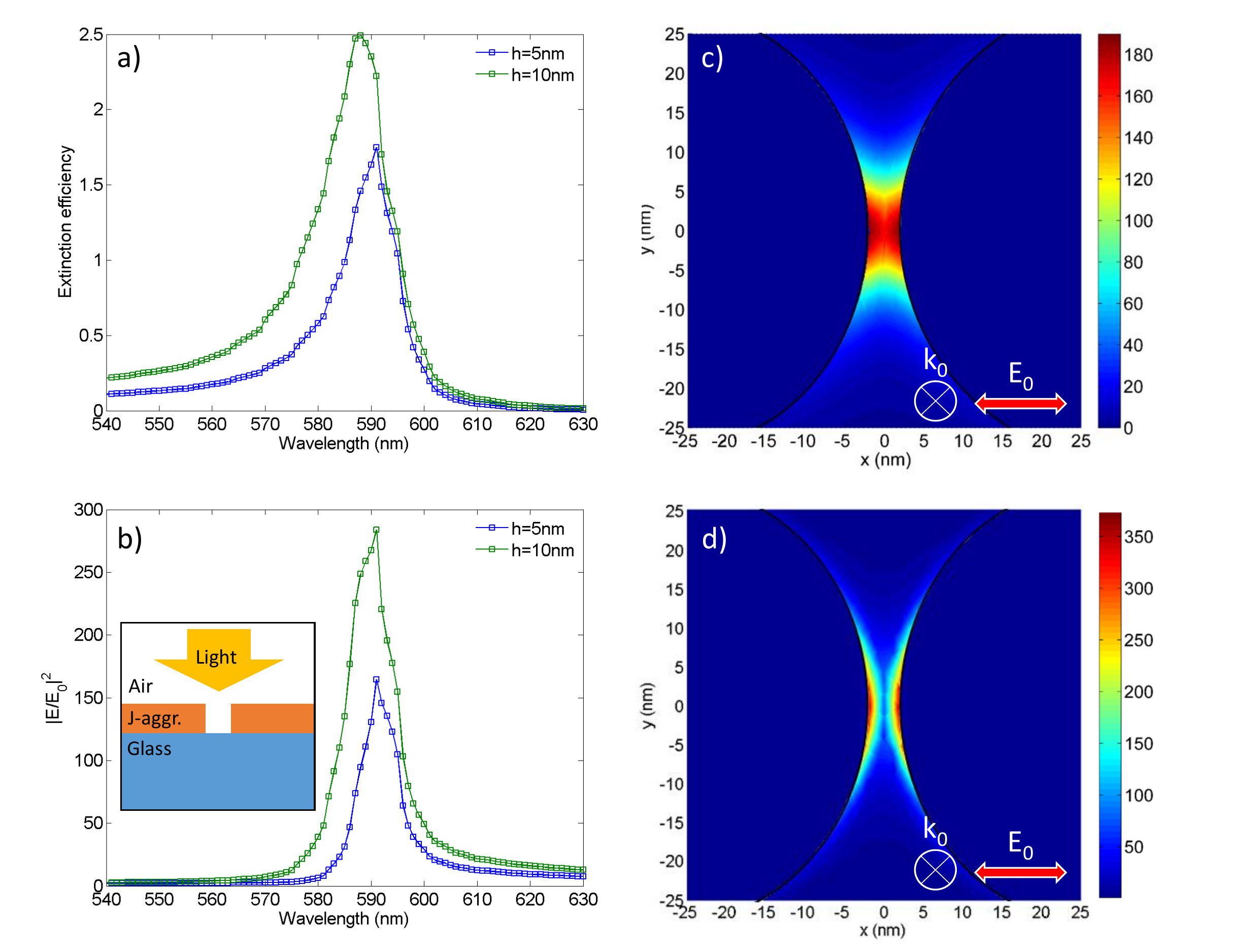}
  \caption{Scattering calculations under plane wave illumination with normal incidence and longitudinal polarization (along the dimer axis)  on TDBC/PDAC nanodisk dimers on a glass substrate. The two disks have a radius $a = 30$ nm and are separated by a gap of $4$ nm. (a) Extinction efficiency for dimers of two different heights $h$. (b) Near-field enhancement spectra calculated at the gap center and at $z =h/2$.
(c) Near-field enhancement distribution around the nanostructure on the plane $z = h/2$ for $h=5$ nm and $\lambda = 591$ nm. (d)
Near-field distribution calculated on the substrate surface  $z = 0$	for $h=5$ nm and $\lambda = 591$ nm.
	}\label{5}
\end{figure}
%=====================================================

It is worth observing that the dielectric function used for all the presented calculations has been obtained from optical data fitted by an isotropic model, thus any optical anisotropy effect in neglected. With J-aggregates having an intrinsically anisotropic structure \cite{Kobayashi1996, Wuerthner2011}, anisotropy of the dielectric response can be expected in thin films as well. At our knoledge there are not specific data on TDBC films, however recently, by using spectroscopic ellipsometry and polarized IR spectroscopy, a strongly anisotropic behavior of thiacarbocyanine dye aggregates films was revealed \cite{Roodenko2013}. It was quantified by means of two dielectric functions: one for the $xy$ plane parallel to the film and the other for the $z$ direction perpendicular to it. These data do not show any  optical anisotropy for polarizations along the xy plane.
In the Supporting Information (Figure SI3) we show optical absorbance measurements at normal incidence for two different orthogonal linear polarizations. The data confirm the absence of optical anisotropy on the $xy$ plane.
All the calculations on nanodisks have been performed by considering input light at normal incidence and hence polarized along the xy plane. In this case the near field generated by the nanodisk along the z axis is very small (beyond one order of magnitude) and thus the presence of a different dielectric function along the z axis does not affect significantly the obtained results. The lack of optical anisotropy in the xy thin film response may be due to the local disorder during the growth process, as evidenced in AFM images in Ref.\ \cite{Bradley2005}.
We also observe that, from available AFM images (see e.g. Ref.\ \cite{Bradley2005}), it is possible to infer some degree of local order in the 100 nm range. Hence it is possible that TDBC nanodisks (or more generally nanostructures) display some degree of optical anisotropy on the xy plane, depending on the specific arrangement of the monomers. In this case the dielectric function that we used for all the calculations can be interpreted as an average quantity. For example for a brickwork type arrangement of the monomers in the aggregates, a larger dipole moment can be expected along the head-tail direction of the aggregates. For the nanodisks here investigated, we would expect an increase of the near-field enhancement for input light polarized along the head-tail direction and a corresponding decrease along the orthogonal direction.
Another interesting issue is the possible dependence of the dielectric function on the size of the nanostructures. As the sharp peak in the dielectric function and the corresponding negative permettivity arises from the interaction of the dye molecules, the dielectric function might strongly depend on the size of the nanostructures. According to accurate theoretical calculations including both the electronic and vibrational degrees of freedom, a chain with 15 monomers fully reproduces the observed shifted and sharp absorption peaks of J-aggregates \cite{Roden2009}. These theoretical analysis is confirmed by femtosecond nonlinear optical experiments on TDBC J-aggregates, showing that the exciton delocalization length at room temperature corresponds to 16 molecules \cite{Burgel1995}. As a consequence, nanostrutures with linear dimensions beyond 30 nm should display the desired optical properties.

Scattering calculations on nanodisk dimers are shown in Fig.\ 5.  The two disks have a radius $a = 30$ nm and are separated by a gap of $4$ nm. We consider dimers of two different heights $h=$ 5, 10 nm.
The extinction efficiency and near-field enhancement spectra are shown in Figs.\ 5a and b under plane wave illumination with normal incidence and longitudinal polarization (along the dimer axis). The enhancement spectra have been calculated at the gap center and $z =h/2$.
The near-field enhancement distribution around the nanostructure on the plane $z = h/2$ under the above described conditions is displayed in Fig.\ 5c.
Figure 5d shows the resulting field distribution calculated on the substrate surface  ($z = 0$). A further concentration of the field around
 the  opposite disks surface can be observed. Analogous calculations based on an excitonic resonance providing a dielectric function with positive real  part \cite{Castriciano2003} displays a negligible near-field enhancement (Supporting Information Figure SI4).

\section{Conclusions}
The presented results demonstrate that J-aggregates nanostructures display LSRs able to confine light  below the diffraction limit, resulting into a strong  near field enhancement comparable  to that induced by gold nanoparticles. 
These results, in view  also of the  huge variety of organic dyes that can aggregate and their chemical flexibility, hold the promise to greatly enlarge the availability of new plasmonic materials with different properties with respect to noble metals. For example J-aggregates display attractive nonlinear optical properties which could be exploited for the realization of switchable LSRs for nanophotonic devices.
Specifically, it has been shown that an ultrafast optical pump is able to significantly affect the dielectric function of J-aggregates \cite{Kobayashi1996,Fofang2011,Kauranen2012,Vasa2013}, hence we expect light-controlled modifications of the LSRs, very sensitive to the real part of the dielectric function, which can be exploited for the realization of ultracompact all-optical switches.
Since analogous molecular aggregates can be found in nature \cite{Coles2014}, acting as antennas in photosynthetic complexes, the results here presented open the intriguing question if photosynthetic organisms \cite{McDermott1995} exploit LSRs to improve light harvesting efficiency \cite{Engel2007,Collini2010}. At the same time J-aggregate nanoantennas could represent a low-cost solution for improving photovoltaic devices  and for the realization of efficient artificial photosynthetic light-harvesting antennas \cite{Kodis2002}.

\section{Methods}
%Scattering calculations have been performed by the finite element method by solving 3D Maxwell's equations. The 3D geometry is discretized using tetrahedral elements. Their size is extremely fine (up to 0.1 nm) expecially around nanodisks and interfaces, in order to better describe the rapid changes in the electromagnetic field. Scattering calculations have been validated by comparing the results with the simulation of the same spherical nanoparticles and dimers studied above with the Mie theory. 

Scattering calculations on spherical nanoparticles and dimers have been performed within the transition-matrix method, initially introduced by Waterman \cite{Waterman1965} as a general technique for computing the e.m. scattering based on the Huygens principle. In the last decades this method has been significantly improved by expanding the incident, the internal and the scattered electromagnetic fields in vector spherical harmonics (VHS) \cite{Borghese2007}. Thanks to the linearity of Maxwell’s equations and  boundary conditions,  when the scatterers, either single or aggregate, are constituted by spherical monomers, the expansion of the electromagnetic fields in terms of VSH allows to analytically relate the multipolar amplitude of incident field to those of the scattered field by means of the T-matrix.  The Transition matrix contains all the information on the microphysical properties of the scatterer, being independent from the state of polarization of incidence field, and from the incident and observation direction. The elements of T-matrix define analytically in far-field, the optical cross section, and in the near-field zone the intensity distribution of the total and scattered field. From the computational point of view, the numerical calculation of the transition matrix requires the inversion of a matrix whose order is, in principle, infinite. However, it is possible to truncate the multipole expansion of the fields, that determine the size of the array, in order to ensure the numerical stability of the results.

Finite element method (FEM) is a differential equation method that computes the scattered time-harmonic electromagnetic field by solving numerically the vector Helmholtz equation subject to boundary conditions at the particle surface. Thanks to the discretization process (mesh construction) of the domains in smaller parts, this technique is suitable to describe systems with complex geometry. The structure, placed on a glass substrate is embedded in a 3D-finite computational domain that is discretized into many small-volume tetrahedral cells. In the far-field zone, at the outer boundary of the finite computational domain, approximate absorbing boundary conditions are imposed. In this way it is possible both to suppress wave reflections back into the domain and to allow the propagation of the numerical outgoing waves as if the domain were infinite \cite{Mittra1990}. In the near-field zone, for the 3D tetrahedral elements, we choose an extremely fine size (up to 0.1 nm) to better describe the rapid changes in the electromagnetic field. By solving 3D Helmholtz's equations together with the boundary conditions in each element of mesh, we can reconstruct the optical behavior of the system, the near-field enhancement distribution around the nanostructures (see figures 4b, 5c-d) and the extinction spectra in far-field. Scattering calculations were validated by comparing the results with the simulation of the same spherical nanoparticles and dimers studied above with the Mie theory.

Conflict of Interest: The authors declare no competing financial interest.

\begin{acknowledgement}
The authors acknowledge financial support from the MPNS COST Action MP1403 Nanoscale Quantum Optics.
The authors also thanks L. Mons\`{u} Scolaro for providing the data on the dielectric function in the Supporting Information and for useful discussions.
\end{acknowledgement}

\begin{suppinfo}

Supporting Information Available: Spontaneous emission enhancement of a TDBC nanosphere;
Scattering calculations based on an excitonic resonance providing a dielectric function with positive real  part for a dimer; Scattering Calculations for a gold nanodisk.

\end{suppinfo}

%========== Bibliography =============
\bibliography{plasticplasmons2}

\end{document}